\pdfoutput=1
\documentclass[12pt]{article}
\usepackage{graphicx}
\usepackage{microtype}
\usepackage{amsmath}
\usepackage[round]{natbib}

\title{Seismic data denoising and deblending using deep learning}
\author{Alan Richardson\footnote{Ausar Geophysical, both authors contributed equally} \and Caelen Feller\footnotemark[\value{footnote}]}

\newcommand{\plot}[3]{
\begin{figure}
\includegraphics{#1.pdf}
\caption{#3}
\label{fig:#1}
\end{figure}}

\begin{document}
\maketitle
\begin{abstract}
An important step of seismic data processing is removing noise, including interference due to simultaneous and blended sources, from the recorded data. Traditional methods are time-consuming to apply as they often require manual choosing of parameters to obtain good results. We use deep learning, with a U-net model incorporating a ResNet architecture pretrained on ImageNet and further trained on synthetic seismic data, to perform this task. The method is applied to common offset gathers, with adjacent offset gathers of the gather being denoised provided as additional input channels. Here we show that this approach leads to a method that removes noise from several datasets recorded in different parts of the world with moderate success. We find that providing three adjacent offset gathers on either side of the gather being denoised is most effective. As this method does not require parameters to be chosen, it is more automated than traditional methods.
\end{abstract}

\section{Introduction}
Seismic data contain noise that must be removed to avoid reducing final image quality. In marine data, this includes cable and swell noise. Traditional methods of noise attenuation include FX deconvolution \citep{gulunay1986fxdecon} and median filtering \citep{liu20081d}. These methods rely on the signal component of the data being coherent and the noise component being incoherent, so a gather domain in which this is true must be chosen. Although generally effective, these methods require human intervention to choose parameters, such as window sizes, and may require multiple iterations in different gather domains. This causes them to have high labor and computational cost.

In order to reduce the duration of seismic data acquisition, simultaneous source or blended acquisition may be used. This occurs when another shot is initiated while arrivals from the previous shot are still being recorded, leading to interference. Removing the interference from shot records separates the data for each shot, allowing processing to continue using regular processing methods. This process is called deblending, and the interference is referred to as blending noise. Deblending methods typically rely on the blending noise being incoherent in domains other than common shot gathers, due to randomness in shot initiation times, while the desired signal remains coherent. Denoising methods may thus be used to remove the blending noise \citep{huo2012simultaneous}. Alternatively, an inversion approach may be used to explicitly separate the data into different shot records that are coherent \citep{moore2008simultaneous}. The inversion approach is usually even more computationally expensive than the denoising methods, and, furthermore, requires knowledge of the initiation time of each shot. The latter issue poses a difficulty in applying the method on old datasets or to remove interference from other nearby surveys being performed simultaneously, where these times may not be available.

Machine learning, in the form of deep neural networks, has recently been used to achieve state of the art results in image denoising tasks \citep{zhang2017beyond}. It is capable of achieving better denoising performance at a lower computational cost than traditional denoising methods, while also requiring less human interaction. The network is trained to identify noise by presenting it with examples of noisy data and the corresponding noise or noise-free data.

Several recent applications of deep learning in seismic data denoising \citep[e.g., ][]{jin2018seismic,liu2018random,yu2018deep,zhang2019random}, and deblending \citep{baardman2019classification,slang2019using}, demonstrate that the benefits of this approach observed in image denoising transfer to seismic data.

Previous seismic applications of deep learning tend to begin training from randomly initialized weights, leading to long training times despite using relatively shallow, custom-designed networks. One could alternatively use deeper networks such as a U-net \citep{ronneberger2015u} that is commonly used in computer vision tasks, and was recently found to be successful for seismic interpolation and denoising by \citet{mandelli2019interpolation}. This not only enables the network to learn more complex operations, potentially leading to better results \citep{simonyan2014very}, but also creates the opportunity to start training from pretrained weights. For popular network architectures that can be used to construct the U-net, such as ResNet34 \citep{he2016deep}, these pretrained weights, trained on tasks such as classifying images in the ImageNet dataset \citep{russakovsky2015imagenet}, are widely available for download. As the initial layers in the network perform basic operations such as identifying edges, the weights often transfer well to other tasks \citep{yosinski2014transferable}.

Another potential shortcoming of some previous proposals that use deep learning in seismic applications is the failure to exploit the high dimensionality of seismic datasets. Many of these proposals only operate on two dimensional gathers of data, ignoring the information contained in adjacent gathers. Figure \ref{fig:neighbour_data} demonstrates the similarity of adjacent common offset gathers, providing information that could be used to more accurately estimate the signal. This is similar to applications of deep learning for video denoising, where the similarity of adjacent frames is exploited by providing several adjacent frames as input when the model is denoising the central frame \citep{claus2019videnn}.

\plot{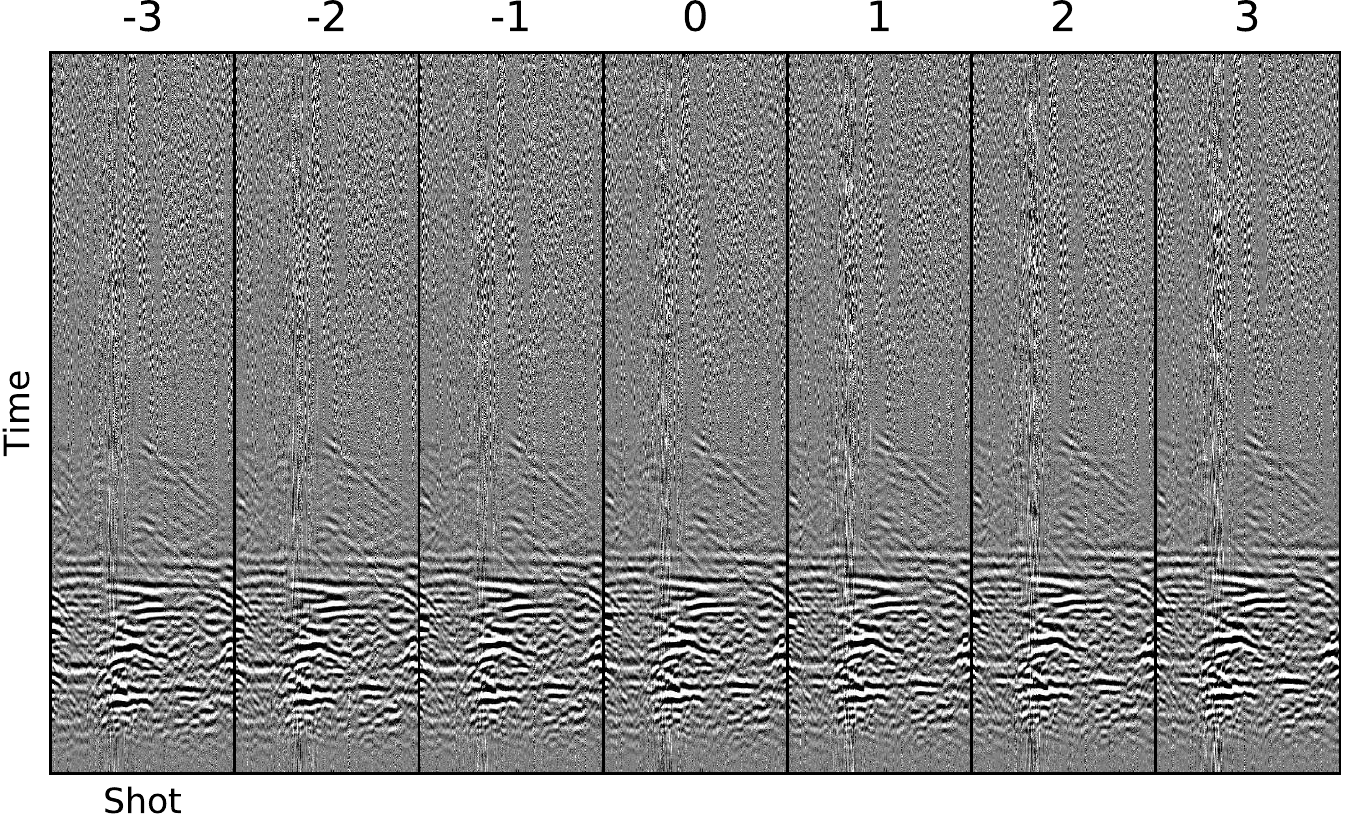}{}{Three adjacent common offset gathers on either side of a central gather exhibit the redundant information contained in higher dimensions of seismic data.}

Finally, several previous proposals were only trained using one dataset. As a result, the network will not have been exposed to all of the variations it may encounter in seismic data, and so its behavior on other datasets is unpredictable. Training using a diverse range of datasets allows the network to experience a wide range of situations and so reduce the risk of overfitting.

Our hypothesis is that a U-net, based on a pretrained ResNet, and provided with information from multiple adjacent gathers, will produce an effective tool for denoising seismic data. Applied to blended data in a gather dimension in which the blending noise is incoherent, it may also be used for deblending. By training it on a diverse training dataset, we believe it will exhibit the ability to generalize well to other datasets not used during training.

\section{Materials and Methods}
\subsection{Network Architecture}
We use a U-net model \citep{ronneberger2015u}, with a ResNet34 architecture as the encoder \citep{he2016deep}, using the implementation in Fast.ai version 1.0.53. The channel dimension is used to provide adjacent common offset gathers as input. ResNet34 expects three input channels. Thus, when the number of adjacent common offset gathers provided on either side of the gather being denoised is not one, we modify the first convolutional layer to have the appropriate kernel shape.

\subsection{Training}

We train the network using synthetic data, which we generate with an acoustic propagator and twenty randomly generated wave speed models. A slice through one of these models is shown in Figure \ref{fig:synthetic_model}. To ensure that the network behaves robustly on unseen data, these wave speed models exhibit more extreme variability than is generally found in reality. We use a Ricker wavelet source with a peak frequency of 35~Hz and a source and receiver spacing of 5~m. Each model is used to create one dataset with 300 towed receivers and 250 shots, with a shot record length of 1.2~s and a time sample interval of 0.001~s. The data are synthetically blended. This is achieved by creating a continuous recording for each dataset, where the shot records from the dataset are added with a time delay between shots. The average time delay and the jitter in the time delay are randomly chosen for each dataset between 0.5 and 1.5~s, and between 0.1 and 0.6~s, respectively. As this time delay is sometimes shorter than the shot record length, the shots may overlap. We then extract overlapping sections of this record corresponding to each shot record, including the blending noise. To increase the ability of the model to recognize other forms of noise, we also add data from real datasets to the synthetic data. This not only adds realistic noise, such as cable and swell noise, but, since the real datasets also contain signal, simulates interference from other nearby surveys. The datasets used are field data from cruises \citet{ew0007}, \citet{ew0207}, and \citet{ew0210}. White Gaussian noise is also added. Example common offset gathers of the input and target synthetic data are shown in Figure \ref{fig:synth_data}. Data from two of the wave speed models are reserved for use as a validation dataset.

\plot{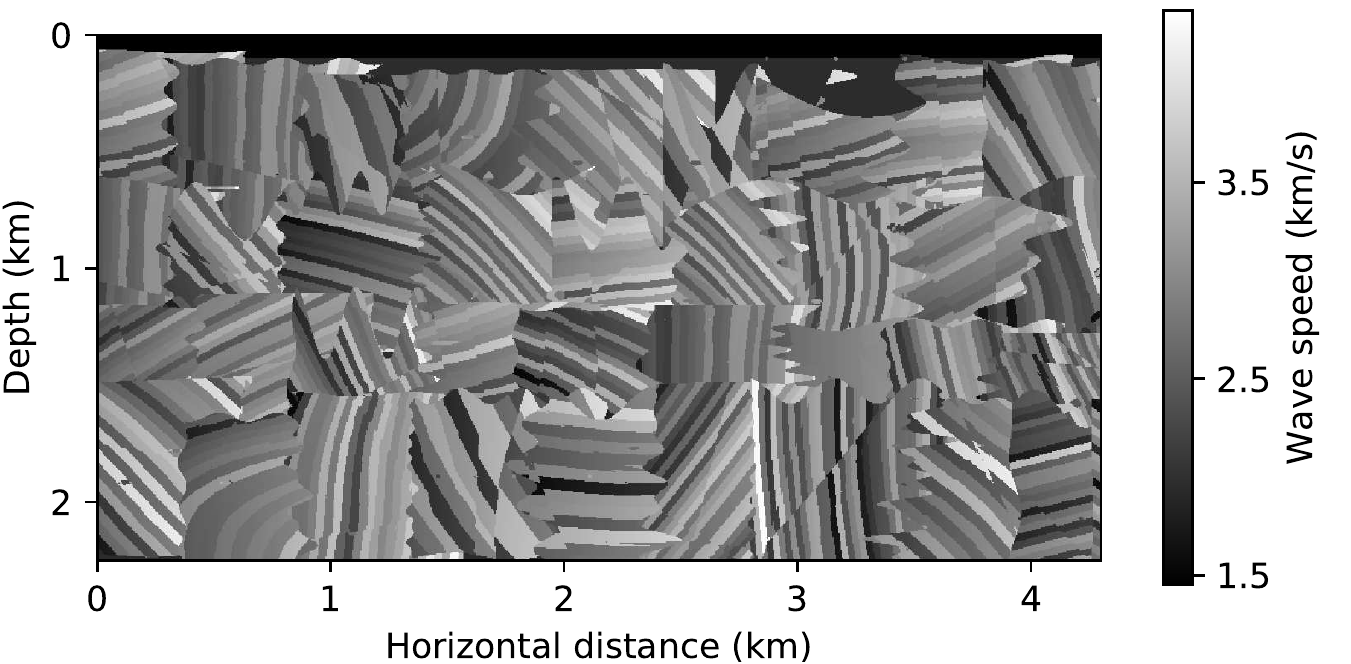}{}{A vertical slice through one of the synthetic models used to generate training data.}
\plot{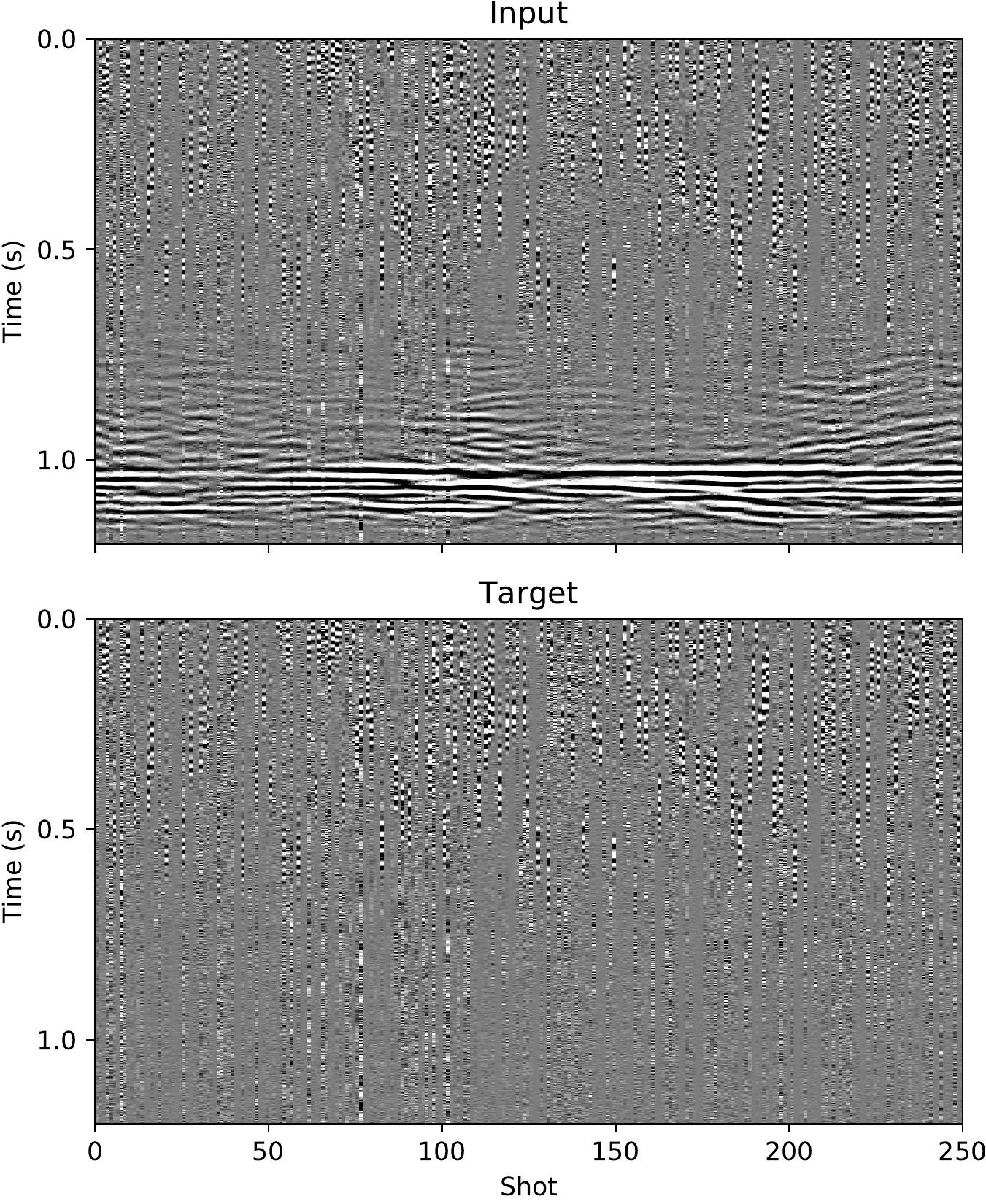}{}{An example noisy common offset gather and the corresponding noise that are used as the input and target during training.}

We initialize the ResNet portion of the model using pretrained weights. These are the weights obtained by training on the ImageNet dataset. We use the default version of these weights distributed with PyTorch \citep{paszke2017automatic}. As these weights are designed for three input channels (the red, green, and blue channels of ImageNet images), we replace the weights in the first convolution with the sum of the pretrained weights in the channel dimension, divided by the new number of input channels, repeated for each new input channel:
\begin{equation}
        W_{c,x,y} = \frac{\sum_{i=1}^3W^p_{i,x,y}}{N},
\end{equation}
where $W$ are the weights used in the first convolutional layer of the network, $W^p$ are the corresponding pretrained weights, $c$, $x$, and $y$ are the channel, vertical, and horizontal indices, respectively, and $N$ is the new number of input channels. The remaining weights in the network (i.e., those not in the ResNet portion) are randomly initialized.

During training, the noisy data are the inputs, and the corresponding noise (the input minus the signal) is the target. We divide each input and target gather by the standard deviation of the input gather. As the pretrained ResNet weights expect the input to have 224 pixels in both height and width, we randomly extract squares of this size from the data. One such square is extracted from each common offset gather in the dataset per epoch. To further increase the range of inputs that the model is exposed to during training, we apply transformations to the data. These convolve the data in time with a random kernel (simulating the use of different source wavelets), randomly set the stride in the time dimension between one and five (altering the apparent frequency of the data), and randomly choose the stride used to select the adjacent common offset gathers between one and four (exposing the model to different receiver spacings).

We use the 1cycle training method \citep{smith2018disciplined} and the AdamW optimizer \citep{loshchilov2017fixing}, with the mean squared error loss function. We initially only train the weights in the model that were randomly initialized, leaving the pretrained weights fixed. We use a learning rate of 0.01 for five epochs. We then train all of the weights for 350 epochs, using Fast.ai's discriminative learning rate that increases from $1\times 10^{-6}$ to $1\times 10^{-3}$ with increasing depth in the model. A batch size of 32 is used throughout training.

\subsection{Test Datasets}
To test the trained model on a variety of inputs, we select four 2D marine streamer datasets from different regions of the world. These are the North Sea, Venezuela, Norway, and New Zealand (Mobil AVO Viking Graben Line 12, and cruises EW0404 \citep{ew0404}, EW0307 \citep{ew0307}, and EW0001 \citep{ew0001}, respectively).

As none of these datasets use blended acquisition, we synthetically blend the data from the North Sea, using the blending procedure described above, with a time delay between shots of $1.8 \pm 0.2$~s. 

To create a common mid-point (CMP) stack image using the North Sea data, we sort the data into CMP gathers, apply normal moveout (NMO) correction, and stack. We use an NMO velocity profile that is constant at 1500~m/s up to 0.5~s, rises to 2000~m/s at 2~s and then remains constant.

The model operates on patches of data with 224 pixels on each side. To apply it to a whole gather, we divide the gather into overlapping patches, apply the model to each, and use the mean value in the overlapping regions. We use a stride of 56 pixels in both time and shot dimensions.

\section{Results}
\subsection{Number of adjacent common offset gathers in input}
We train the network using, as additional channels in the input, varying numbers of adjacent common offset gathers on both sides of the gather that corresponds to the target. Each configuration is trained using the training procedure described above, but with only 10 epochs in the second training stage. The training is repeated 45 times for each configuration, with different random initializations, and the minimum mean squared error loss achieved on the validation dataset is recorded.

The results, displayed in Figure \ref{fig:neighbour_comp}, show that the loss decreases rapidly from 0 to 1 adjacent gathers, indicating that the model uses the information contained in these additional channels effectively. The loss continues to decrease up to 3 adjacent gathers, before increasing slightly and then remaining almost constant up to 8 adjacent gathers. 

\plot{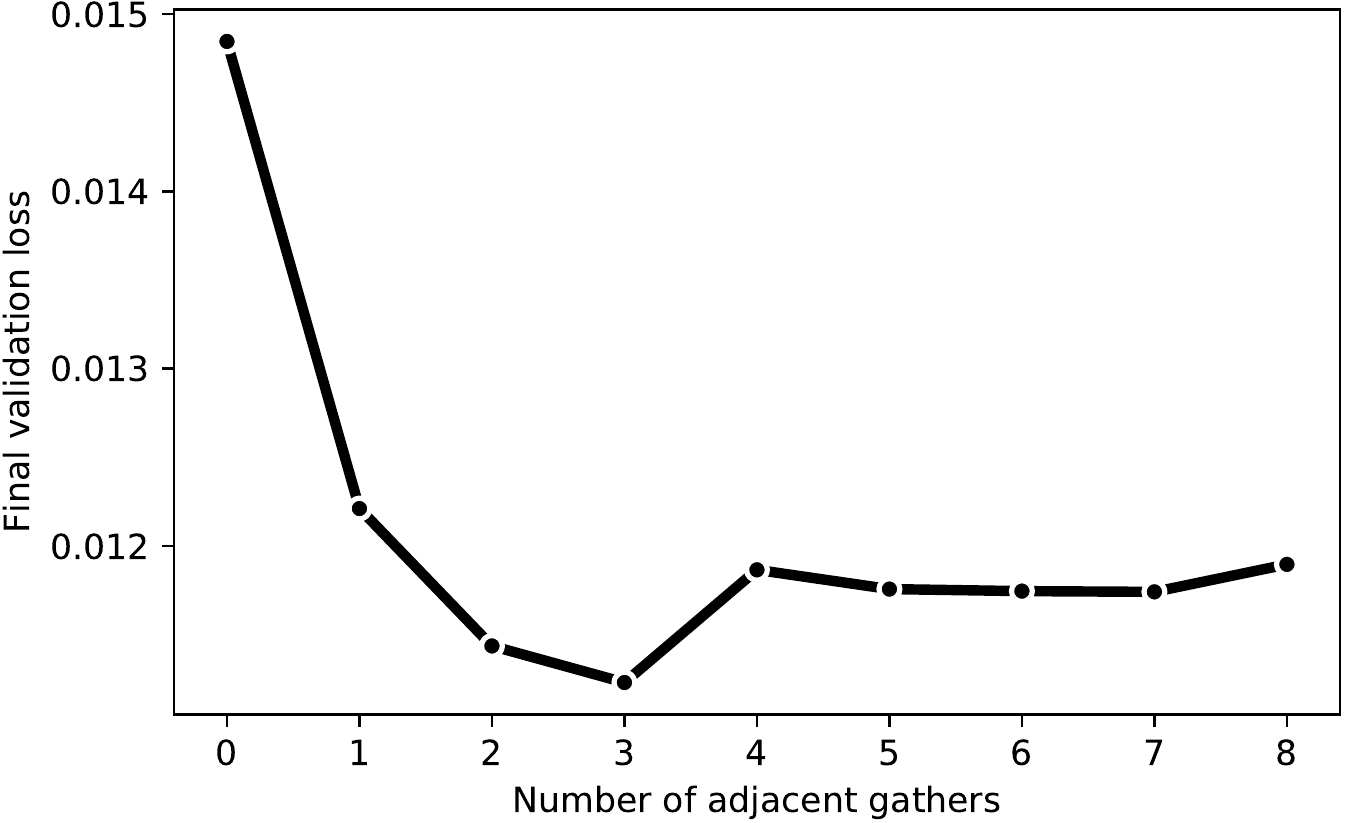}{}{The best results are obtained when three adjacent common offset gathers on both sides of the gather being denoised are provided in the input.}

Subsequent results use 3 adjacent gathers.

\subsection{Deblending}
We apply the trained network to the blended North Sea data. Figure \ref{fig:vg_gathers} shows the result on one common offset gather. We calculate the mean signal-to-noise ratio using
\begin{equation}
\text{SNR}(data) = \frac{1}{n_s}\sum_{s=1}^{n_s} 10\log_{10}\left(\frac{||\mathit{signal}_s||^2_2}{||\mathit{signal}_s - \mathit{data}_s||^2_2}\right),
\end{equation}
where $\mathit{signal}$ is the unblended dataset, $s$ is the shot index, and $n_s$ is the number of shots in the dataset. Using this metric, the SNR of the input data is -3.6~dB, while that of the output deblended data is 16.5~dB. This is comparable with improvements in SNR reported using other techniques on different datasets, such as \citet{chen2018deblending} reporting an increase in SNR from 1.62~dB to 16.55~dB, and \citet{zu2018hybrid} reporting an increase from 0~dB to 19.18~dB.

\plot{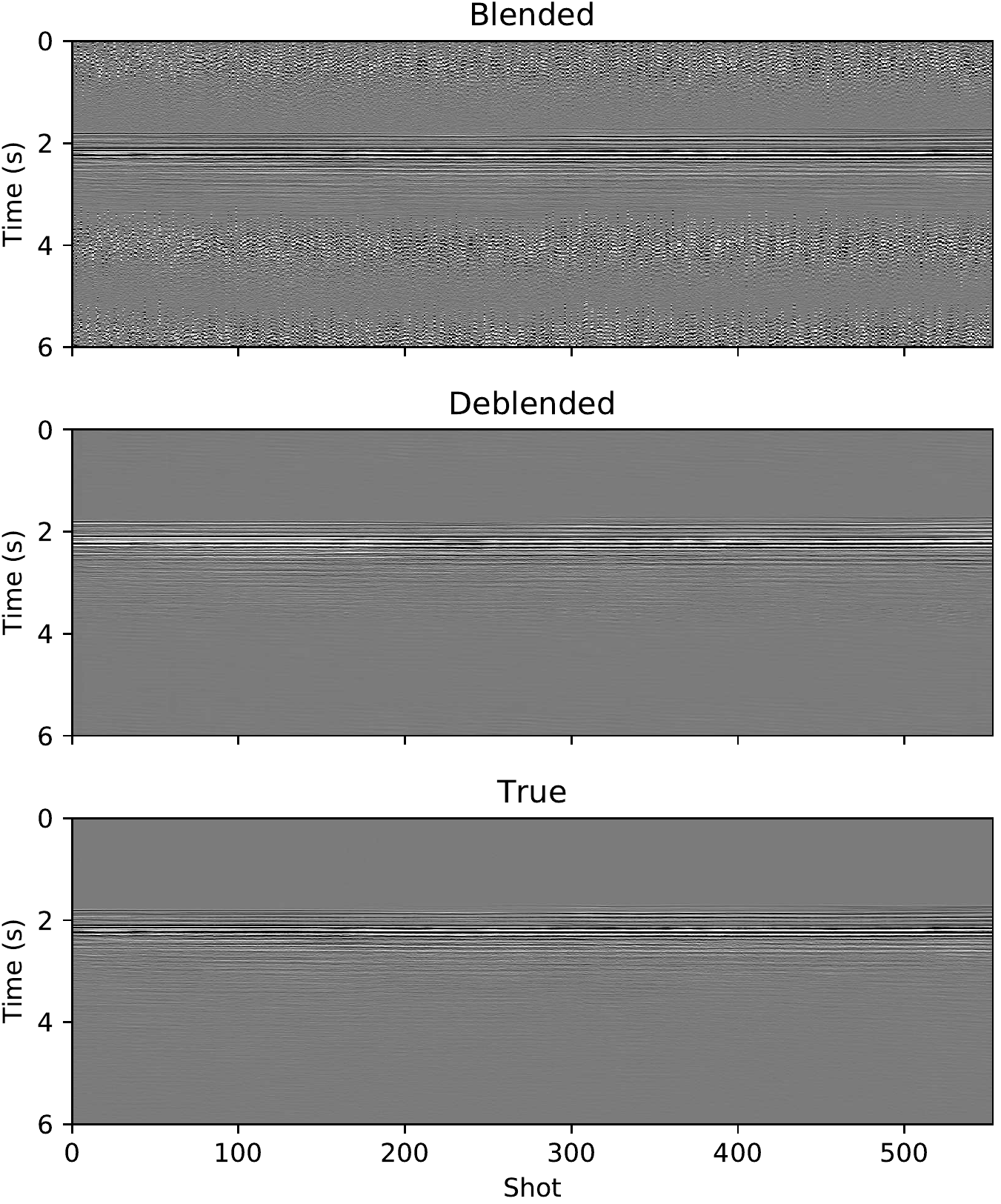}{}{Common offset gathers before (input) and after deblending, and the unblended data.}
\plot{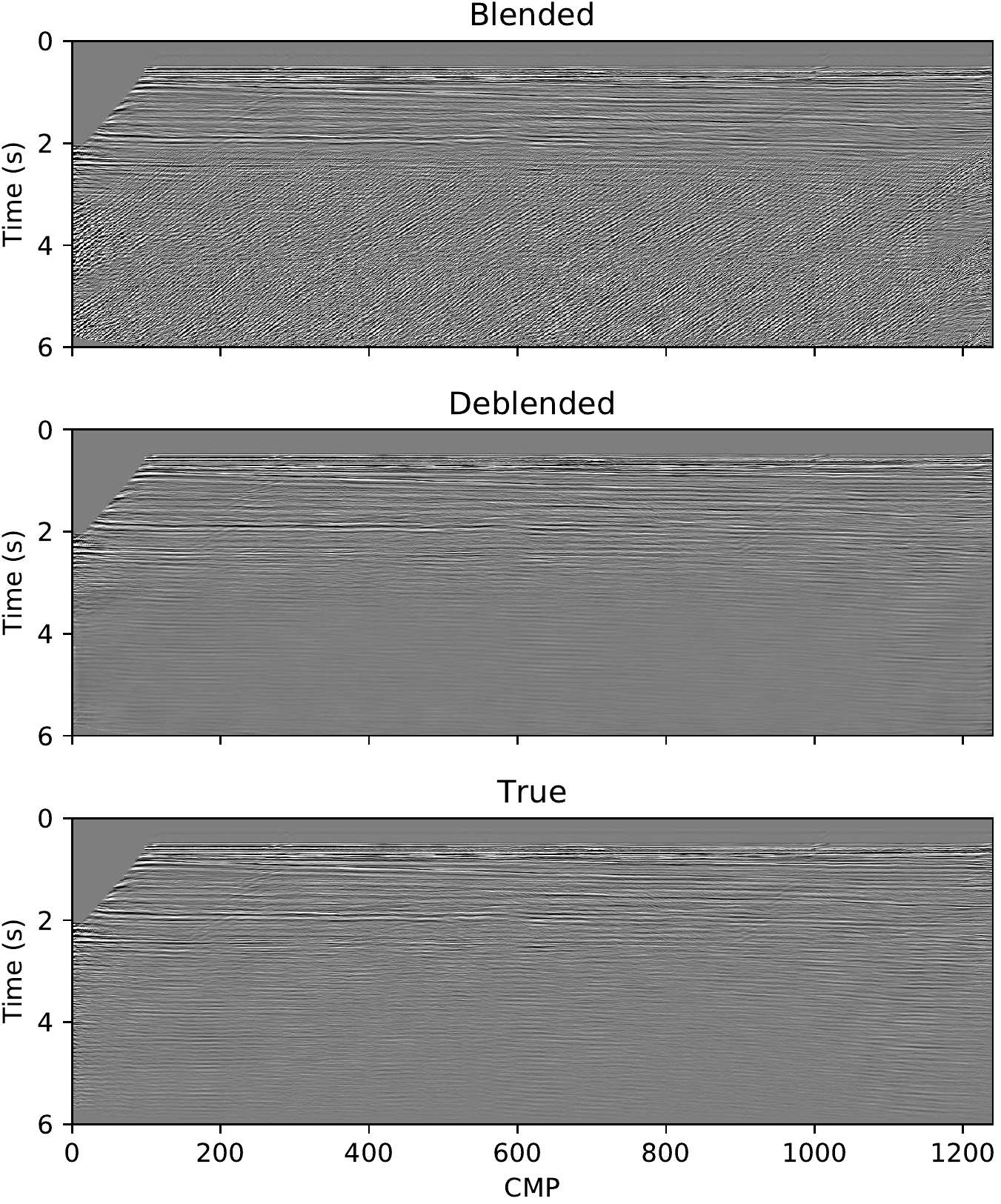}{}{CMP stack images produced using the North Sea data, before (input) and after deblending, and using the unblended data. Deblending removes the blending noise artifacts. The underlying signal is reduced, but still visible.}

We create CMP stack images to produce the result shown in Figure \ref{fig:vg_stacks}. The blending artifacts are less severe in the output.

\subsection{Noise removal}
When we apply the model, without any modification, to three noisy datasets from different parts of the world, we obtain the results shown in Figure \ref{fig:other_data}. The model effectively removes the noise. Some signal is also visible in the estimated noise.
\plot{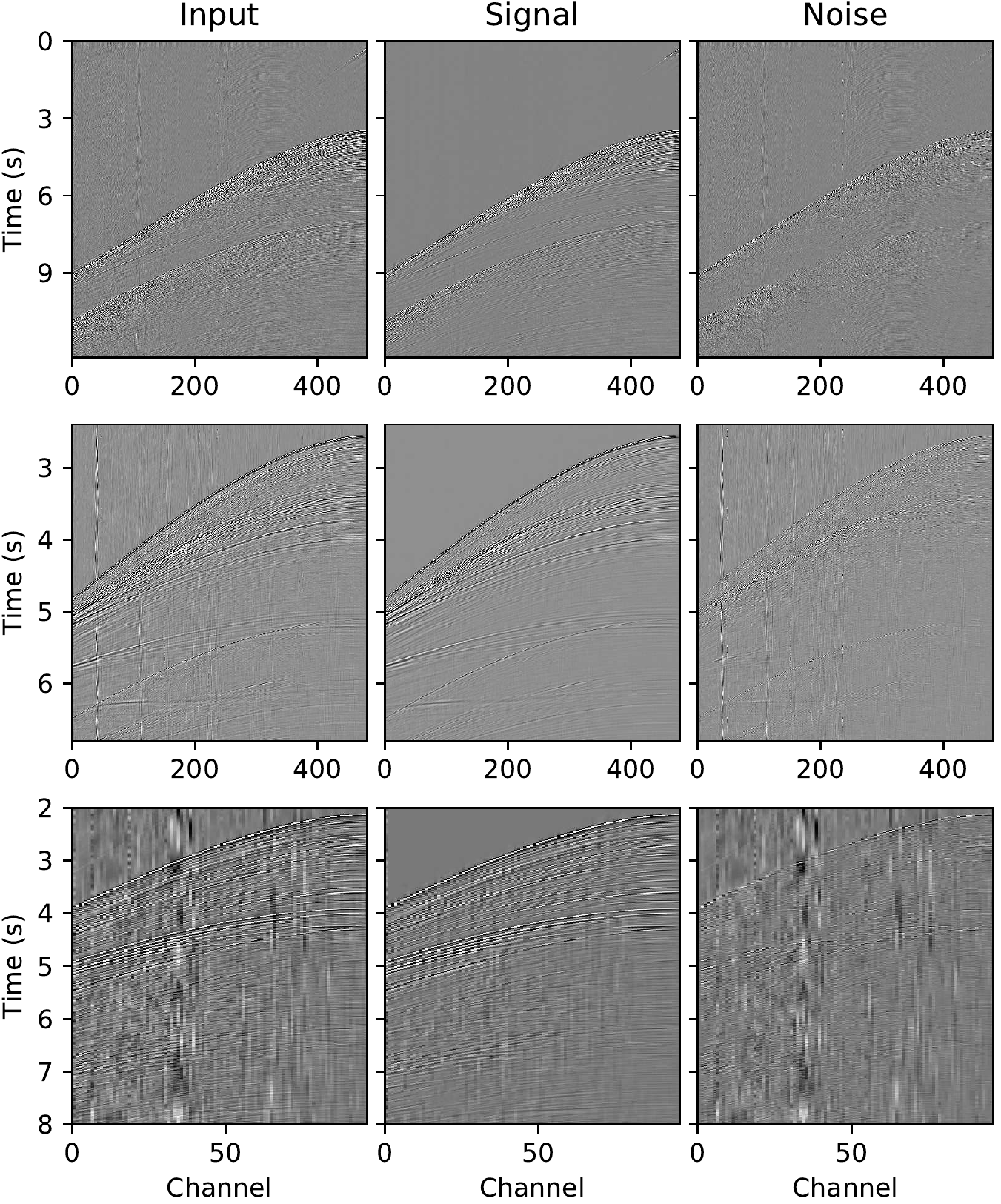}{}{Noise removal from marine data acquired in Venezuela, Norway, and New Zealand. The output of the network, the estimated noise, is shown on the right, while the estimated signal, shown in the center, is obtained by subtracting the estimated noise from the input. The majority of the noise is removed, but some signal is also lost.}

\section{Discussion}
\paragraph{Increasing loss with number of adjacent gathers} In theory, the loss should not increase as the number of adjacent gathers provided as input to the model increases, as the model may disregard the additional gathers if they do not provide useful information. It is possible that the loss increases in our experiment when more than three adjacent gathers are provided as beyond this point the useful information provided by additional gathers is not sufficient to compensate for the larger number of training epochs that will be required due to the increasing number of weights that need to be trained.

\paragraph{Alternative adjacent gathers} We use adjacent common offset gathers as the additional input channels provided to the network. Alternatives include constructing the additional input channels so that slices of the input in the channel dimension correspond to NMO corrected CMP gathers. In this setup, the blending noise is incoherent in two dimensions of the input, while the signal should be coherent and relatively flat. This may aid the network to remove the blending noise, but results in higher computational cost, and also requires a velocity model to perform the NMO correction.

\paragraph{Signal loss} The apparent signal leakage into the estimated noise, visible when the model is applied to real datasets, is undesirable. Inspection of the results suggests that small variations in amplitude are being identified by the model as noise. This is likely to be due to the model being trained using synthetic data, where such variations are not present. Incorporation of some real data into the training dataset may help to reduce this effect.

\paragraph{Application to land data} Denoising methods that rely on identifying signal by its coherency are less reliable for seismic data acquired on land. This is because static shifts, caused by near-surface variations, and irregular acquisition spacing due to obstacles, may cause the signal to also be incoherent across traces. Applying static corrections and interpolation to regularize the acquisition geometry may thus be necessary before attempting to attenuate the noise.

\paragraph{Black box method} Deep learning-based methods may be regarded as a black box, where it is not possible to explain why particular results were produced or to predict how the methods will respond to new inputs. The compensation for this sacrifice is that the methods are capable of more sophisticated operations than we are able to express in mathematically derived approaches. A median filter is easily understood, but will struggle when the underlying signal is not well approximated by a line and when there are crossing events. In contrast, a neural network can, in theory, learn to recognize these situations and handle them appropriately.

\section{Conclusion}
After being trained once, the method denoised and deblended datasets from different parts of the world that it had not seen previously, with moderate accuracy. The method removed noise effectively, but also removed some signal. The hypothesis is therefore validated for situations where the speed of the method is more important than preserving all of the signal, such as during fast-track processing and quality control.

\bibliography{paper}{}
\bibliographystyle{plainnat}
\end{document}